\documentclass[journal]{IEEEtran}

\usepackage{algorithm}
\usepackage{algorithmicx}
\usepackage{algpseudocode}
\usepackage{CJK}
\usepackage[top=2cm, bottom=2cm, left=2cm, right=2cm]{geometry}

\usepackage{amssymb}

\usepackage{graphicx}
\usepackage{subfigure} 
\usepackage{array}
\usepackage{amsmath}
\usepackage{amsmath,bm}
\usepackage{booktabs}
\usepackage{makecell}
\usepackage{multirow}
\usepackage{stfloats}
\usepackage{diagbox}
\usepackage{cite}
\usepackage[numbers,sort&compress]{natbib}
\usepackage{threeparttable}
\usepackage{varwidth}
\usepackage{stfloats}
\usepackage{graphicx}
\usepackage{soul}
\ifCLASSINFOpdf
\else
\fi
\usepackage{amsmath}
\begin{document}
\title{Sensor-Movement-Robust Angle Estimation for 3-DoF Lower Limb Joints Without Calibration }

\author{Chunzhi~Yi,
       Feng~Jiang$^*$,
        Zhiyuan~Chen,
        Baichun~Wei,
        Hao~Guo,
        Xunfeng~Yin,
        Fangzhuo~Li,
        and~Chifu~Yang % <-this % stops a space
\thanks{C.Yi, H.Guo and C.Yang are with the School
of Mechatronics Engineering, Harbin Institute of Technology, Harbin,
Heilongjiang, 150001 China e-mail: chunzhiyi123@gmail.com(C.Yi).}% <-this % stops a space
\thanks{Z.Chen is with the School of Computer Science, University of Nottingham, Malaysia}% <-this % stops a space
\thanks{X.Feng and F.Li are with the School of Mechanical Engineering,Harbin Engineering University, Harbin, Heilongjiang, China.}% <-this % stops a space
\thanks{B.Wei and F.Jiang are with the School of Computer, Harbin Institute of Technology, Harbin, Heilongjiang, 150001 China and Pengcheng Laboratory, 
Shenzhen, Guangdong ， China e-mail: fjiang@hit.edu.cn(F.Jiang).}}%% <-this % stops a space
\maketitle

\begin{abstract}
Inertial measurement unit (IMU)-based 3-DoF angle estimation methods for lower limb joints have been studied for decades, however the calibration motions and/or careful sensor placement are still necessary due to challenges of real-time application. This study proposes a novel sensor-movement-robust 3-DoF method for lower-limb joint angle estimation without calibration. A realtime optimization process, which is based on a feedback iteration progress to identify three joint axes of a 3-DoF joint, has been presented with a reference frame calibration algorithm, and a safe-guarded strategy is proposed to detect and compensate for the errors caused by sensor movements. The experimental results obtained from a 3-DoF gimbal and ten healthy subjects demonstrate a promising performance on 3-DoF angle estimation. Specially, the experiments on ten subjects are performed with three gait modes and a 2-min level walking. The root mean square error is below 2 deg for level walking and 5 deg for other two gait modes. The result of 2-min level walking shows our algorithm’s stability under long run. The robustness against sensor movement are demonstrated through data from multiple sets of IMUs. In addition, results from the 3-DoF gimbal indicate that the accuracy of 3-DoF angle estimation could be improved by 84.9$\%$ with our reference frame calibration algorithm. In conclusion, our study proposes and validates a  sensor-movement-robust 3-DoF angle estimation for lower-limb joints based on IMU. To the best of our knowledge, our approach is the first experimental implementation of IMU-based 3-DoF angle estimation for lower-limb joints without calibration.
\end{abstract}

%
%

% \fi
\IEEEpeerreviewmaketitle

\section{Introduction}
\IEEEPARstart{T}{he} need for real-time human motion measuring in pathological human movement analysis \cite{chen2012location}, stability evaluation of locomotion \cite{dingwell2007differences}, virtual reality systems \cite{wittmann2015assessment}, and human-robot interaction \cite{ding2016imu} has driven researchers to develop novel tracking techniques. Indoor motion capture systems (e.g. optical motion capture systems\cite{lamine2017evaluation} and magnetic resonance systems based on imaging methods\cite{graichen2005effect},\cite{de2006six}), which utilize image processing techniques, are accurate enough to be a gold standard. However, it requires controlled laboratory settings, trained staff and costly facilities, such fatal flaws limit their use in real-time application scenarios.

For real-time application, inertial measurement units (IMU) with multi-axis gyroscopes, accelerometers and magnetometers were widely used to estimate hip, knee and ankle angles \cite{liu2009ambulatory}, \cite{favre2006new} and \cite{picerno2008joint}.  To estimate 3-Degree-of-Freedom (3-DoF) angles for lower-limb joints, the IMU-based angle estimation algorithm can be decoupled to two steps. The first step is to estimate the absolute orientation of IMUs. The data measured by an IMU are described as local vectors in a sensor-fixed frame, $[s_i]$. Herein, orientation tracking tehniques were developed to transform the vectors from $[s_i]$ to the earth frame in order to describe all the measurements in the same coordinate frame. Having obtained the absolute orientations of the IMUs placed on two adjacent segments, the second step is to infer joint angles by developing body-fixed frames using biomechanical constraints.

%\cite{roetenberg2005compensation,roetenberg2007estimating,yun2006design,luinge2004inclination} 

In this research,  in order to develop a novel sensor-movement-robust based 3-DoF method for lower limb joint angle estimation without calibration, challenges in  both  steps  need to be solved. Firstly, due to the diverse characteristics of measurements from each IMU, the absolute orientations of IMUs estimated by orientation tracking techniques are actually described in different reference frames \cite{brennan2011quantification},\cite{vitali2017method}, rather than in the earth frame. This will definitely lead to a large error in the resulted angle estimation. Currently used methods for calibrating reference frames suffer from either a linear approximation of "the time-varying deviation" between reference frames \cite{brennan2011quantification} or a rough calibration metrics \cite{vitali2017method}. Thus, a calibration method, which is able of providing a more accurate compensation for the reference frame deviation needs to be developed with comprehensive metrics. In addition, most works of the sensor-to-body alignment for 3-DoF lower-limb joints depend on functional calibration postures.  Other than being cumbersome, such postures could result in additional errors if subjects cannot perform the postures accurately. Thus, it is necessary to develop a method for estimating 3-DoF joint angles without using functional calibration postures. Moreover, when muscles, on which the IMUs are mounted, oscillate severely, or even when sensors are moved with respect to their mounted segments, how to detect the movements and re-align IMUs is still a challenge under real-application scenarios. Online detection and safeguarded strategy against sensor movements should be included in the algorithm.

Aiming at solving the problems above, we developed a sensor-movement-robust algorithm for estimating 3-DoF angles lower-limb joints without using calibration postures.  The main contributions of this paper are summarized as follow:
\begin{itemize}

\item To the best of our knowledge, this is the first study to realize a realtime detection and correction for sensor movements  during the progress of estimating lower limb joint angles.

\item A novel 3-DoF geometric constraint of lower-limb joints was proposed , with which joint axes can be further estimated without employing functional calibration postures.

\item A pointwise calibration method, which utilized the fused calibrations of magnetic field and gravity vectors, was proposed to effectively overcome the problem of time-varying deviation between reference frames and ensure the accuracy of the whole algorithm.

\end{itemize}

 This paper has been organized as follow:  Related works are presented in Section  II. Section III details the methodology and experiments have been explained in Section  IV. Results and discussions of the entire research work have been given in Section V. Section VI concludes the whole study.

\section{Related Work}

Although the IMU-based angle estimation technique has been studied for decades, there are still severe challenges for estimating 3-DoF lower-limb joint angles for real-time application. One challenge  is that the second step of currently used techniques, developing body-fixed frames, still suffer from either a set of cumbersome predefined functional calibration procedures or carefully aligned body-to-sensor relationship usually associated with body landmarks defined by the International Society of Biomechanics (ISB) recommendations \cite{wu2002isb}. In \cite{liu2009ambulatory} and \cite{favre2006new}, the coordinate axes of sensor-fixed frames were assumed to be collinear with joint axes, such an assumption makes the accuracy heavily depend on IMUs alignment. Picerno et al.\cite{picerno2008joint} proposed an analytical-based calibration method using a calibration device to distinguish body landmarks. Similarly, several calibration procedures were proposed to transform sensors’ measurement into orientations of bone-embedded anatomical frames\cite{cappozzo1996position,besier2003repeatability,charlton2004repeatability,hagemeister2005reproducible}. However, it was reported that the repeatability of analytical-based calibration methods was obviously worse than functional calibration\cite{schache2006defining,mannel2004establishment}. 

Other works \cite{o2007inertial,palermo2014experimental,roetenberg2009xsens} proposed various functional calibration procedures , in which subjects were asked to perform a set of pre-determined tasks to define each orientation of biological axes with respect to sensor-fixed frames. Functional calibration presented robustness towards the execution of calibration movements\cite{besier2003repeatability} and the load applied to the joint\cite{mannel2004establishment,marin2003correction}. A functional calibration with passive movements was introduced by Favre et al.\cite{favre2009functional} in order to improve robustness. Recently, Valencia et al.\cite{vargas2016imu} proposed an IMU-to-body alignment method where an initial upright posture needs to be performed by subjects to align a coordinate axis with gravity. But some additional strategies, such as accurate sensor placement or standing along the direction of geomagnetic field, still need to be adopted to align other two coordinate axes.  

In order to avoid the need for such calibration procedures for 3-DoF joint angle estimation, some related work in estimating angles of joints with fewer DoFs does give encouragement. Seel et al.\cite{seel2014imu} proposed a 1-DoF joint angle estimation method for knee and ankle, which fused the integration of angular rate along joint axis and the inclination of accerlation based on exploring the kinematic constraints of a hinge joint.  In \cite{muller2016alignment}, Muller et al. extended Seel's work towards 2-DoF joints, in which the absolute orientations of IMUs were required. 

It should be noted that among all the previous work in estimating joint angles for lower-limb joints, none of the published work was proven to be robust against the sensor movements after calibration procedures. Although Muller and Seel's work were reported to be robust against skin artificial movements, no published work have experimentally validated how to detect and compensate for large sensor movements.

Another challenge is that  the accuracy of estimating IMUs' absolute orientation has huge effect to the performance of angle estimation. The error of estimating IMUs' absolute orientations with motion tracking algorithms will make the vectors described in sensor-fixed frames into reference frames rather than the earth frame. The reference frames of IMUs mounted on different segments vary due to IMUs' individual signal corruptions. Brennan et al.\cite{brennan2011quantification} validated to what extent the absolute orientation estimation error could lead to the accuracy reduction when estimating 3-DoF angle. Therefore, in our research work, in order to reduce the error caused by the deviation between the two reference frames, a reference frame calibration algorithm should be included in 3-DoF joint angle estimation. Favre et al.\cite{favre2008ambulatory} defined gravity as the Z axis of each reference frame. The deviation between two reference frames was calibrated by uniforming the projected angular rates of each IMU in the horizontal plane while the subjects were to perform a hip abduction and adduction.  However, due to the changing of magnetic distortion and movement-caused acceleration, the deviation between two reference frames is time-varying. The calibration in this work cannot be updated over time because of the calibration posture. Based on Favre's single calibration method, Brennan et al.\cite{brennan2011quantification} performed the calibration procedure before and after data collection in order to linearly interpolate the calibration angle by time.  The accuracy represented by the root mean square errors indicated relatively large errors of such approximation. Vitali et al. \cite{vitali2017method} proposed a calibration algorithm for estimating 3-DoF knee angle to dynamically calculate a "correction" direction cosine matrix, which assumes that the flexion/extension axis estimated by Seel's 1-DoF joint angle estimation method \cite{seel2014imu} should be the same in reference frames. Although Vatali's work provides some encouragement, the 1-DoF joint axis estimated by \cite{seel2014imu} is based on the hinge-joint approximation, which is not acccurate enough on 3-DoF lower-limb joints. Thus, a comprehensive metrics of deviation between the reference frames should be given.

\setlength{\parskip}{0em}

Motivated by the challenges from the literature review, we presented how accurately 3-DoF angle of lower-limb joints can be estimated without performing pre-defined postures or careful alignment. A novel numerical optimization-based estimation strategy has been proposed where an algorithm was embedded to detect sensor movements and compensate the consequent errors.  In our experiments, IMU signals recorded from ten healthy subjects and a 3-DoF gimbal were used to respectively evaluate the potential of the designed 3-DoF angle estimation algorithm and the errors from each step of our algorithm have been analyzed separately.  The promising results of this study will aid the future development of IMU-based motion measuring for gait-related application. 

 

\section{Methodology}
In this  research, a sensor-movement-free estimation of the orientation relationship between two segments  beside  a  3-DoF  joint  without  introducing  calibration movements has been proposed. More specifically, a point-wise reference frame calibration combining with an orientation tracking algorithm is firstly used to estimate IMUs' absolute orientations in the same reference frame. Then  the biological axes of lower-limb joints are estimated through geometric constraints of each joint to develop alignment between body-fixed frames and sensor-fixed frames.  In this section, the principle of our algorithm is presented separately in  \ref{SecA} and \ref{SecB}. The implementation is concretely described in \ref{SecC}.

\subsection{Calibration for Two Reference Frames} 
\label{SecA}

Firstly, the absolute orientation of each IMU is calculated by an improved complementary filter proposed in our previous work\cite{yi2018estimating} to dynamically fuse readings from gyroscope, accelerometer and magnetometer. 

As stated above, due to the different data characteristics of signals of the two IMUs mounted on each segment, the absolute orientations of both IMUs are described in two different reference frames $[g_1]$ and $[g_2]$, which can be represented by quaternions $q_{s_1}^{g_1}$, $q_{s_2}^{g_2}$. To compensate for errors caused by the difference between two reference frames, a common reference frame should be employed, which is namely to estimate a correction quaternion from $[g_2]$ to $[g_1]$. 

%\begin{figure}[ht]
%\centering
%\includegraphics[width = 5cm]{01.png}
%\caption{The schematic diagram of reference frame calibration}
%\label{fig:01}
%\end{figure}

Without including movement-caused acceleration, the acceleration vectors, $\bm{a_1}$ and $\bm{a_2}$, should be identical in the common reference frame, which contributes to the construction of the correction quaternion, given by:
\begin{align}
\nonumber q_{acc}=\cos(\theta_{acc}/2) \cdot \begin{bmatrix} 1 \\ 0 \\ 0 \\ 0 \end{bmatrix}+\sin(\theta_{acc}/2) \cdot \begin{bmatrix} 0 \\ \bm{W_{acc}} \end{bmatrix} 
\end{align}
\begin{align}
\bm{W_{acc}}=({q_{s_1}^{g_1}} \otimes {\bm{a_1}}) \times ({q_{s_2}^{g_2}} \otimes {\bm{a_2}}) 
\end{align}
\begin{align}
\label{corr_acc}
\nonumber \theta_{acc}=\frac {{q_{s_1}^{g_1}} \otimes {\bm{a_1}}\times {q_{s_2}^{g_2}} \otimes {\bm{a_2}}}{\Vert{({q_{s_1}^{g_1}} \otimes {\bm{a_1}}) \times ({q_{s_2}^{g_2}} \otimes {\bm{a_2}})}\Vert}
\end{align}
Where $\bm{W_{acc}}$ is the rotation axis, $\theta_{acc}$ is the rotation angle and $q_{acc}$ is the correction quaternion calculated from gravity. Similarly, the identical magnetic field vectors $\bm{m_1}$ and $\bm{m_2}$ without magnetic distortion could contribute to another correction quaternion, given by:
\begin{align}
\nonumber q_{mag}&=\cos(\theta_{mag}/2)\cdot \begin{bmatrix} 1 \\ 0 \\ 0 \\ 0 \end{bmatrix}+\sin(\theta_{mag}/2) \cdot \begin{bmatrix} 0 \\ \bm{W_{mag}} \end{bmatrix}
\end{align}
\begin{align}
\bm{W_{mag}}&=({q_{s_1}^{g_1}} \otimes {\bm{m_1}}) \times ({q_{s_2}^{g_2}} \otimes {\bm{m_2}})
\end{align}
\begin{align}
\label{corr_mag}
\nonumber \theta_{mag}&=\frac {{q_{s_1}^{g_1}} \otimes {\bm{m_1}}\times {q_{s_2}^{g_2}} \otimes {\bm{m_2}}}{\Vert{({q_{s_1}^{g_1}} \otimes {\bm{m_1}}) \times ({q_{s_2}^{g_2}} \otimes {\bm{m_2}})}\Vert}
\end{align}

However, due to the corruption of acceleration and magnetometer readings, neither of the correction quaternions calculated above are accurate enough to represent the rotational relationship between $[g_1]$ and $[g_2]$. Herein, a weighted sum of these two correction quaternions is used to make a fusion.
\begin{align}
q_{corr}=k_{mag} \cdot q_{mag}+k_{acc} \cdot q_{acc}
\end{align}

To calculate the angle of a 3-DoF joint, the detailed description of geometric constraint will be presented, which will start from the 1-DoF joint and then extending to 3-DoF joint. 

\subsection{Sensor-to-Body Alignment Based on Geometric Constraints } \label{section:2.2}
\label{SecB}
%\IEEEpubidadjcol

\subsubsection{Geometric Constraints of a 1-DoF Joint}

For lower-limb joints, there is always a main axis around which the joint rotates the most time during gait cycles. The main axis usually corresponds to flexion/extension for hip and knee, and dorsiflexion/plantarflexion for ankle. Simplifying a lower-limb joint as a hinge joint, the main axis could be estimated through its geometric constraint based on the method introduced by Seel et al.\cite{seel2014imu}, which is given by:
\begin{align}
\label{e4}
\Vert{\bm{\omega_1} \times \bm{j_1^{2D}}}\Vert-\Vert{\bm{\omega_2} \times \bm{j_2^{2D}}}\Vert=0
\end{align}
Where $\bm{\omega_1}$, $\bm{\omega_2}$ are angular rate vectors of each segment with respect to the sensor-fixed frames $[s_1]$ and $[s_2]$ respectively, and $\bm{j_1^{2D}}$ , $\bm{j_2^{2D}}$ are the same unit joint axis but described in $[s_1]$ and $[s_2]$ respectively. 
%\begin{figure}[ht]
%\centering
%\includegraphics[width = 6cm]{02.png}
%\caption{The joint axis vector of a hinge joint}
%\label{fig:02}
%\end{figure}

According to \cite{seel2014imu}, the geometric constraint of joint position vectors can be exploited as:
\begin{align}
\label{e6}
\nonumber \Vert{\bm{a_1}-\bm{\dot{\omega_1}} \times \bm{o_1}-\bm{\omega_1} \times (\bm{\omega_1} \times \bm{o_1})}\Vert\\
-\Vert{\bm{a_2}-\bm{\dot{\omega_2}} \times \bm{o_2}-\bm{\omega_2} \times (\bm{\omega_2} \times \bm{o_2})}\Vert=0
\end{align}
Where $\bm{a_1}$ , $\bm{a_2}$ and $\bm{\omega_1}$ , $\bm{\omega_2}$ are readings of accelerometers and gyroscopes mounted on each segment, respectively. $\bm{o_1}$ and $\bm{o_2}$ are vectors from the origin of each sensor frame to the rotating center.
%\begin{figure}[ht]
%\centering
%\includegraphics[width = 6cm]{03.png}
%\caption{The joint position vector of a 3-DoF joint}
%\label{fig:03}
%\end{figure}

%Having obtained vectors of joint axis and sensor position, the rotation matrix between sensor-fixed frame and %body-fixed frame could be calculated. Compared with the fusion method presented in \cite{seel2014imu}, the %kinematics calculation method directly fuse $\bm{j_1^{2D}}$ and $\bm{j_2^{2D}}$, $\bm{o_1}$ and $\bm{o_2}$ %without manully tuning parameters. 

%\begin{align}
%\label{e7}
%\nonumber R_A^B&=\begin{bmatrix} \cos(\bm{e_{xA}} \cdot \bm{e_{xB}})&\cos(\bm{e_{yA}} \cdot %\bm{e_{xB}})&\cos(\bm{e_{zA}} \cdot \bm{e_{xB}})\\
%\cos(\bm{e_{xA}} \cdot \bm{e_{yB}})&\cos(\bm{e_{yA}} \cdot \bm{e_{yB}})&\cos(\bm{e_{zA}} \cdot %\bm{e_{yB}})\\
%\cos(\bm{e_{xA}} \cdot \bm{e_{zB}})&\cos(\bm{e_{yA}} \cdot \bm{e_{zB}})&\cos(\bm{e_{zA}} \cdot %\bm{e_{zB}})\\ \end{bmatrix}\\
%&=\begin{bmatrix}\bm{e_{xA}^{[B]}}&\bm{e_{yA}^{[B]}}&\bm{e_{zA}^{[B]}}\end{bmatrix}
%\end{align}

According to the construction of rotation matrix,  $R_{s_i}^{b_i}$ that represents the rotation matrix between body-fixed frame [$b_i$] and sensor-fixed frame [$s_i$] is defined by the $\bm{x_{s_i}^{b_i}}$ coincident with the joint axis, which is assumed to be vertical to the sagittal plane.
\begin{align}
\label{xsi}
\bm{x_{s_i}^{b_i}}=\bm{j_i^{2D}} , i=1,2 
\end{align}
\begin{align}
\label{zsi}
\bm{z_{s_i}^{b_i}}=\frac{\bm{x_{s_i}^{b_i}} \times \bm{o_i}}{\Vert{\bm{x_{si}^{bi}} \times \bm{o_i}}\Vert} , \bm{y_{s_i}^{b_i}}=\bm{z_{s_i}^{b_i}} \times \bm{x_{s_i}^{b_i}}
\end{align}

\begin{align}
\label{rsi}
R_{si}^{bi}=\begin{bmatrix}\bm{x_{s_i}^{b_i}}&\bm{y_{s_i}^{b_i}}&\bm{z_{si}^{b_i}}\end{bmatrix}
\end{align}
\subsubsection{Geometric Constraints of a 3-DoF Joint}
Estimating the main axis through the method presented above, although straightforward, suffers from the perturbation caused by rotation around other axes, which will consequently lead to relatively large errors and even divergence. Besides, angles of other DoFs are also of importance for gait-related researches. Herein, a general condition is considered to develop geometric constraints of a 3-DoF joint as an extension of the abovementioned method. Considering two segments connected by a 3-DoF joint, 
%\begin{figure}[h]
%\centering
%\includegraphics[width = 7.5cm]{04.png}
%\caption{Three joint axes of a 3-DoF joint}
%\label{fig:04}
%\end{figure}
%\begin{figure*}[htbp]
%\centering
%\includegraphics[width = 10cm]{05.png}
%\caption{The flow diagram of the whole 3-D angle estimating algorithm}
%\label{fig:05}
%\end{figure*}
the rotation of a segment relative to the other one can be decoupled into three sequent rotations around three axes. As shown in Fig. S1 \footnote{It should be noted that all the figures and tables named as S$i$ are presented in the Supplementary information.}, $\bm{j_1^{3D}}$ and $\bm{j_2^{3D}}$ denote the axes fixed on each segment respectively, while $\bm{j_3^{3D}}$ denotes the main axis. In order to obtain a further relaxed constraint on 3-DoF lower-limb joint, the assumptions are organized as follow:
\begin{itemize}
\item{The main axis, $\bm{j_3^{3D}}$, is perpendicular to the other two axes, which meets the defination of ISB} \cite{wu2002isb}.

\item As adopted by \cite{muller2016alignment}, either of the other two axes $\bm{j_1^{3D}}$ and $\bm{j_2^{3D}}$ prossesses a fixed relative orientation with its corresponded segment (i.e. $\bm{j_1^{3D}}$ - segment 1, $\bm{j_2^{3D}}$ - segment 2). If sensor movements are not considered, the coordinates of  $\bm{j_1^{3D}}$ and $\bm{j_2^{3D}}$ are fixed in $[s_1]$ and  $[s_2]$ respectively (i.e. $\bm{j_1^{3D}}$ - $[s_1]$, $\bm{j_2^{3D}}$ - $[s_2]$).
\end{itemize}

 To unify the description of these axes, a transitional coordinate frame, the reference frame [$g$], should be introduced to describe all the data from both IMUs mounted on both segments. The relationship between angular rates of two segments is given by:
\begin{align}
\label{e11}
[\bm{\omega_2}]_g-[\bm{\omega_1}]_g=\omega_{j_1}[\bm{j_1^{3D}}]_g+\omega_{j_2}[\bm{j_2^{3D}}]_g+\omega_{j_3}[\bm{j_3^{3D}}]_g
\end{align}
Where $\bm{\omega_1}$, $\bm{\omega_2}$ represent two vectors of angular rate described in local sensor-fixed frames, ${\omega_{j_1}}$, ${\omega_{j_2}}$ and ${\omega_{j_3}}$ represent scalar angular rates of each joint axis, $\bm{j_1^{3D}}$, $\bm{j_2^{3D}}$ and $\bm{j_3^{3D}}$ represent three unit joint axes described in sensor-fixed frames, $[\quad]_g$ denotes the description of a vector in the reference frame. Multiplying $[\bm{j_1^{3D}}]_g$ and $[\bm{j_2^{3D}}]_g$ on both sides of the equation, equation (\ref{e11}) could be transformed into:

\begin{align}
\label{e12}
\nonumber f(\bm{j_1^{3D}} , \bm{j_2^{3D}})=([\bm{\omega_2}]_g-[\bm{\omega_1}]_g) \cdot ([\bm{j_1^{3D}}]_g \times [\bm{j_2^{3D}}]_g)  \\
-\omega_{j3}[\bm{j_3^{3D}}]_g \cdot ([\bm{j_1^{3D}}]_g \times [\bm{j_2^{3D}}]_g)=0
\end{align}

One of the joint axes $\bm{j_2^{3D}}$ is known as the same vector as the main axis, $\bm{j_1^{2D}}$ and $\bm{j_2^{2D}}$ . Given that $\bm{j_3^{3D}}$ is perpendicular to the other two axes, the scalar angular rate $\omega_{j3}$ is equal to the projection of $[\bm{\omega_2}]_g-[\bm{\omega_1}]_g$ on this axis, which is:
\begin{align}
\label{e13}
\omega_{j3}=([\bm{\omega_2}]_g-[\bm{\omega_1}]_g) \cdot [\bm{j_3^{3D}}]_g
\end{align}

It could be seen that estimating $\bm{j_1^{3D}}$, $\bm{j_2^{3D}}$ and $\bm{j_3^{3D}}$ has no need for a specific placement, which can be very useful when orientations of IMUs towards segments are unknown. After obtaining the coordinates of $\bm{j_1^{3D}}$ and $\bm{j_2^{3D}}$ in sensor-fixed frames, a sensor-to-body alignment could be achieved by determining the orientation of body-fixed frames.
\begin{align}
\label{e14}
\nonumber\theta_{s_1}^{b_1}&=\arccos({\begin{bmatrix}1&0&0\end{bmatrix}}' , \bm{j_1^{3D}})\\
\nonumber q_{s_1}^{b_1}&=\begin{bmatrix}\cos\Big(\frac{\theta_{s_1}^{b_1}}{2}\Big)\\
\sin\Big(\frac{\theta_{s_1}^{b_1}}{2}\Big) \cdot ({\begin{bmatrix}1&0&0\end{bmatrix}}' , \bm{j_1^{3D}}) \end{bmatrix}\\
\theta_{s_2}^{b_2}&=\arccos({\begin{bmatrix}0&0&1\end{bmatrix}}' , \bm{j_2^{3D}})\\
\nonumber q_{s_1}^{b_1}&=\begin{bmatrix}\cos\Big(\frac{\theta_{s_2}^{b_2}}{2}\Big)\\
\nonumber \sin\Big(\frac{\theta_{s_2}^{b_2}}{2}\Big) \cdot ({\begin{bmatrix}0&0&1\end{bmatrix}}' , \bm{j_2^{3D}}) \end{bmatrix}
\end{align}
\subsection{Algorithm Implementation}

\label{SecC}

Based on the product of the rotation matrixes and quaternions calculated above, the orientation relationship between two body-fixed frames can be estimated. As shown in Fig. S2, some details still need to be implemented in order to complete the whole algorithm. In this section, details will be discussed by explaining each part of the proposed algorithm.

\subsubsection{The Feedback-based Iteration Progress of Calculating 3-DoF Joint Axes}
\begin{figure}[h]
\centering
\includegraphics[width = 8cm]{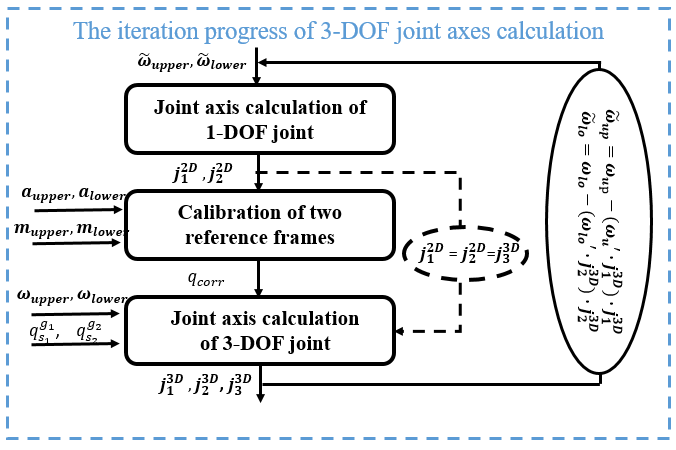}
\caption{The diagram of feedback-based iteration progress}
\label{fig:06}
\end{figure}

Considering the existence of main rotation in each lower-limb joint, the accuracy of 3-DoF angle estimation can be improved by applying geometric constraints of 1 DoF joint and 3-DoF joint iteratively. A complete iteration is presented in Fig. \ref{fig:06}. The main axis described in each sensor-fixed frame, simplifying a 3-DoF joint as a hinge joint, is estimated using 1-DoF joint geometric constraints. Then, plugging the estimated main axis as the known third axis into equation (\ref{e12}) and (\ref{e13}), the geometric constraint of a 3-DoF joint, the other two axes described in each sensor-fixed frame, $\bm{j_1^{3D}}$ and $\bm{j_2^{3D}}$ , can be calculated to represent the rotation axes of abduction and inner rotation. 

It is shown in Fig. \ref{fig:06} that angular rates, $\bm{\omega_{up}}$ and $\bm{\omega_{lo}}$ , measured by each IMU beside the joint are subtracted by their projection on each axis respectively to reduce the errors of the main axis estimation caused by additional rotations, given by:

\begin{align}
\label{angular}
\nonumber \bm{\omega_{up}} = \bm{\omega_{up}} - (\bm{\omega_{up}}' \cdot \bm{j_1^{3D}}) \cdot\bm{\omega_{up}} \\
\bm{\omega_{lo}} = \bm{\omega_{lo}} - (\bm{\omega_{lo}}' \cdot \bm{j_2^{3D}}) \cdot\bm{\omega_{lo}}
\end{align}

 The updating of angular rates constructs a feedback, which contributes to an iteration progress. With the increasing iteration times, the accuracy of estimation could be consequently improved, however, with the price of increasing computing time. Given that the tradeoff between computing efficiency and accuracy, the number of iteration is limited to 6 in order to maintain an acceptable accuracy.

\subsubsection{Numerical Optimization Methods}
As stated in section \ref{section:2.2}, the geometric constraints of anatomical axes are described as a problem of minimizing cost functions.  
%\begin{align}
%\label{e15}
%F(x)=\frac{1}{2} \sum_{i=1}^n f_i(x)
%\end{align}
Utilizing the Gauss-Newton method as presented in \cite{seel2014imu}, the coordinates of $\bm{j_1^{2D}}$ and $\bm{j_2^{2D}}$, $\bm{o_1}$ and $\bm{o_2}$ could be obtained within several iterations. Regarding $\bm{j_3^{3D}}$ as the same vector as $\bm{j_1^{2D}}$ or $\bm{j_2^{2D}}$, equation (\ref{e12}), working as the cost function, could be used to calculate the coordinates of other two axes $\bm{j_1^{3D}}$ and $\bm{j_2^{3D}}$. However, unlike the calculation of $\bm{j_1^{2D}}$ and $\bm{j_2^{2D}}$, Jacobian of derivative-based methods is not straightforward enough to be constructed because $\bm{j_1^{2D}}$ and $\bm{j_2^{2D}}$ can also be treated as a function of  $\bm{j_1^{3D}}$ and $\bm{j_2^{3D}}$. To this end, a secant version of the Levenberg-Marquardt method is applied in this scenario \cite{madsen1999methods}. 
%\begin{align}
%\frac{\partial f_i}{\partial x_j}(\bm{x})=\frac{f_i(\bm{x}+\delta\bm{e_j})-f_i(\bm{x})}{\delta}
%\end{align}
Following this secant version, the coordinates of $\bm{j_1^{3D}}$ and $\bm{j_2^{3D}}$ could be calculated by minimizing cost function $f(\bm{j_1^{3D}}, \bm{j_2^{3D}})$. 

In addition to joint axis calculation, a calibration of two reference frames, combined with $q_{s_1}^{g_1}$ and $q_{s_2}^{g_2}$ estimation, is also embedded in the iteration progress shown in Fig. \ref{fig:06}. The performance of $q_{corr}$ is determined by its two fusion coefficients $k_{mag}$ and $k_{acc}$. For quantification purposes, the deviation of $[\bm{j_1^{2D}}]_{{g_1}}$ and $[\bm{j_2^{2D}}]_{{g_1}}$ is introduced here to evaluate to what extent coordinates in  $[g_2]$ is rotated by $q_{corr}$ into those in $[g_1]$, given by: 
\begin{align}
\label{corr}
\nonumber [\bm{j_1^{2D}}]_{{g_1}}&=q_{s_1}^{g_1} \otimes \bm{j_1^{2D}}
\end{align}
\begin{align}
[\bm{j_2^{2D}}]_{{g_1}}&=q_{s_2}^{g_2} \otimes q_{corr} \otimes \bm{j_2^{2D}}
\end{align}

Regarding $f_i(k_{mag},k_{acc})=\Vert [\bm{j_1^{2D}}]_{{g_1}}-[\bm{j_2^{2D}}]_{{g_1}} \Vert$ as the cost function, fusion coeficients, $k_{mag}$ and $k_{acc}$, can be estimated through the secant version of L-M method. Hence, the calibration of reference frames synthesizes the information from local magnetic field, gravity and the main axis of the 3-DoF joint, which results in a comprehensive measuring. 

 \begin{figure}[htbp]
\centering
\includegraphics[width = 4.5cm]{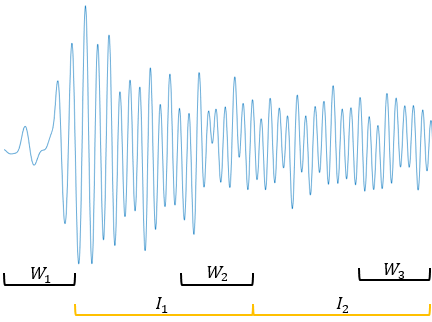}
\caption{Data windowing scheme}
\label{fig:07}
\end{figure}

For the sake of real-time calculation, a sliding window $(W_n)$ is chosen to segment the measurements from each sensor such that the axes' coordinates could be estimated for the last $N$ measurements. As shown in Fig. \ref{fig:07}, two sliding windows are separated by an interval $(I_n)$ within which the coordinates of joint axes estimated in the last sliding window are used to calculate joint angles.

\subsubsection{Detection and Safeguarded Strategy against Sensor Movement}
No published work presented how to detect and re-align IMUs (sensor-fixed frames) to body-fixed frames when IMUs are moved with respect to the segments they are mounted on. Traditionally, the only solution to sensor movements is to terminate the data collection and perform calibration procedures again. Seel and Muller's algorithms, developed for estimating 1-DoF angle and 2-DoF angle respectively, could gradually update the coordinates of anatomical axes based on sliding windows. However, the measurements before sensor movements could no longer be used to estimate axes and angles. So a gradual update is not practical enough especially when the estimated angles are used as input to other applications.

In order to solve this problem, a comparison between outputs of each iteration is presented to ensure the variety of $\bm{j_1^{3D}}$, $\bm{j_2^{3D}}$ and $\bm{j_3^{3D}}$ is under a boundary. A metric of such variety is given by:
\begin{align}
\label{V}
V=\sum_{i=1}^{3}v_i\Vert(\bm{j_i^{3D}})_{iter}-(\bm{j_i^{3D}})_{det}\Vert
\end{align}
Where $(\bm{j_i^{3D}})_{iter}$ denotes the coordinates of each axis output by the iteration progress, $(\bm{j_i^{3D}})_{det}$ denotes the axes estimated for detection, $v_i$ is the weight for the $i$th axis. Because the main axis varies slowly  under normal conditions, its weight $v_3$ is set to $0.2$ while $v_1$ and $v_2$ are set to $0.4$ respectively. The thresholds for judging sensor movements are 1.5 for hip and knee, 2.0 for ankle. During an interval $I_n$, estimation algorithm still works for estimating $(\bm{j_i^{3D}})_{det}$ with 50 measurements. When $V$ is detected to be larger than a threshold, the calculation during this interval will be terminated. And a new sliding window will then be initiated to update estimates of axes, during which the axes $(\bm{j_i^{3D}})_{det}$ are used to estimate angles.

\begin{figure}[htbp]
\centering
\includegraphics[width = 6cm]{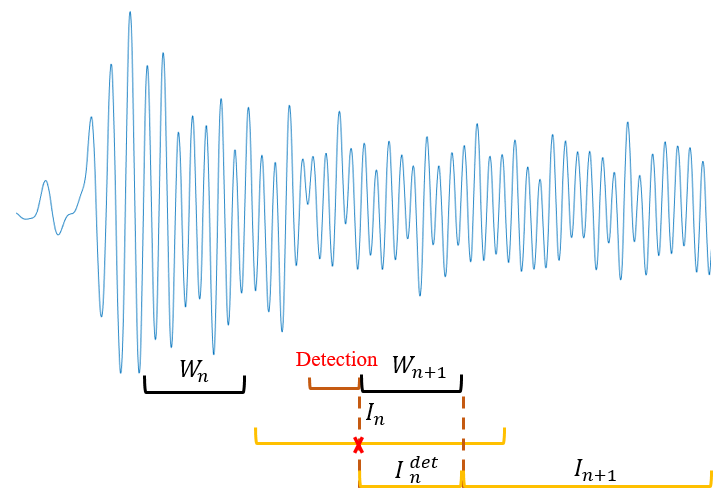}
\caption{The detection and safeguarded strategy for sensor movement}
\label{fig:08}
\end{figure}

\subsubsection{Angle estimation}
Based on the three axes estimated by the whole iteration, orientation relationship $R_{b_1}^{b_2}$ between two segments (i.e. two body-fixed frames) can be constructed. As for a 1-DoF joint, the rotational matrix $R_{b_1}^{b_2}$ can be available by multiplying the matrixes representing the orientations of each frame. 
\begin{align}
\label{rb1b2}
R_{b_1}^{b_2}=R_{b_1}^{s_1} \cdot R_{s_1}^{g_1} \cdot (R_{g_2}^{g_1})^{-1} \cdot  R_{g_2}^{s_2} \cdot R_{s_2}^{b_2} 
\end{align}
Where $R_{b_1}^{s_1}$, $R_{g_2}^{s_2}$ are the inverse of $R_{s_1}^{b_1}$ and $R_{s_2}^{g_2}$ respectively, $R_{g_2}^{g_1}$ is the rotation matrix conversed from $q_{corr}$. 
%\begin{align}
%\nonumber R_{b_1}^{s_1}=(R_{s_1}^{b_1})^{-1}=(R_{s_1}^{b_1})'\\
%R_{g_2}^{s_2}=(R_{s_2}^{g_2})^{-1}=(R_{s_2}^{g_2})'
%\end{align}
%\begin{align*}
%R_{g_2}^{s_2}&=R(q_{corr})\\
%&=\begin{bmatrix}q_0^2+q_1^2-q_2^2-q_3^2&2q_1q_2+2q_0q_3&2q_1q_3-2q_0q_2\\
%2q_1q_2-2q_0q_3&q_0^2-q_1^2+q_2^2-q_3^2&2q_2q_3+2q_0q_1\\
%2q_1q_3+2q_0q_2&2q_2q_3-2q_0q_1&q_0^2-q_1^2-q_2^2+q_3^2 \end{bmatrix}
%\end{align*}
Transforming $R_{b_1}^{b_2}$ into a unit quaternion $q_{b_1}^{b_2}$, the angle around the main axis $\angle j_3$, can be calculated as:
\begin{align}
\label{angj3}
\angle j_3=2 \cdot \cos^{-1}(q_{b_1}^{b_2}(1))
\end{align}

For a 3-DoF joint, the core step of calculating inner rotation and inversion angles is to eliminate the rotation around the main axis. The calculation of such angles depends on decoupling $q_{b_1}^{b_2}$ into Euler angles.During the decoupling progress, the rotation around $\bm{j_3^{3D}}$ will be decoupled into additional rotations around $\bm{j_1^{3D}}$ and $\bm{j_2^{3D}}$ due to the discordance of $\bm{j_3^{3D}}$ and the sencond rotation axis of Euler-angle decoupling. To eliminate such additional rotations, a quaternion that represents the rotation around $\bm{j_3^{3D}}$ could be constructed in reference frame $[g_1]$. 
\begin{align}
\label{qj3}
q_{j_3}=\begin{bmatrix}\cos\big(\frac{\angle j_3}{2}) \\ \sin\big(\frac{\angle j_3}{2}) \cdot [\bm{j_3^{3D}}]_{g_1} \end{bmatrix}
\end{align}

By multiply the inverse of $q_{j_3}$, a pseudo orientation relationship between body-fixed frames, ${\tilde{q}}_{b_1}^{b_2}$ can be described as:
\begin{align}
\label{e22}
%\thinmuskip=0mu
{\tilde{q}}_{b_1}^{b_2}=(q_{s_1}^{b_1})^{-1}\otimes q_{s_1}^{g_1}\otimes q_{corr}\otimes(q_{j_3})^{-1}\otimes (q_{s_2}^{g_2})^{-1}\otimes q_{s_2}^{b_2}
\end{align}

Thus, according to equation (\ref{e14}), the angles around $\bm{j_1^{3D}}$ and $\bm{j_2^{3D}}$ could be estimated by decoupling ${\tilde{q}}_{b_1}^{b_2} $ into a sequential rotation of $X$ and $Z$ axes.

\floatname{algorithm}{Algorithm}
\renewcommand{\algorithmicrequire}{\textbf{Input:}}
\renewcommand{\algorithmicensure}{\textbf{Output:}}

\begin{algorithm}
\caption{3-DoF angle estimation for lower-limb joints}
\begin{algorithmic}[1] %每行显示行号
\Require $\bm{\omega_1}, \bm{\omega_2}, \bm{a_1}, \bm{a_2}, \bm{m_1},\bm{m_2}$
\Ensure $\angle j_1$,  $\angle j_2$,  $\angle j_3$, $\bm{j_1^{3D}}$, $\bm{j_2^{3D}}$, $\bm{j_3^{3D}}$, $\bm{o_1}$, $\bm{o_2}$
\State initialize $\bm{j_1^{3D}}$, $\bm{j_2^{3D}}$, $\bm{j_3^{3D}}$, $\bm{o_1}$, $\bm{o_2}$
\While {$sliding$ $windows$}
\For{$i = 0 \to 6$}
\State subtracting angular rates using Eq. (\ref{angular})
%: $\bm{\omega_{up}}, \bm{\omega_{lo}}$ 
\State estimating the main axis using Eq. (\ref{e4})
%: $\bm{j_1^{2D}}$,  $\bm{j_2^{2D}}$
\State calibrating reference frames using Eq. (\ref{corr_acc}), (\ref{corr_mag}), (\ref{corr})
%\Return $q_{corr}$
\State obtaining $\omega_{j3}$ using Eq. (\ref{e13}) and $\bm{[j_3^{3D}]_{g}} =  \bm{[j_1^{2D}]_{g}}$
\State estimating 3-DoF joint axes using Eq. (\ref{e12})
\EndFor
\State \Return $\bm{[j_1^{3D}]_{g_1}}$, $\bm{[j_2^{3D}]_{g_1}}$ and $\bm{[j_3^{3D}]_{g_1}}$
\EndWhile
\While {$intervals$}
\State $result$ and $\bm{[j_i^{3D}]_{det}}$ $\gets$ \Call{SenMove}{$\bm{\omega_{up}},\bm{\omega_{lo}}$}
\If {$result == 1 $ }
\State \textbf{STOP} $intervals$ 
\State \textbf{do} step {$2 - 11$}
\State \textbf{do} step ${16 - 23 }$ using $\bm{[j_i^{3D}]_{det}}$
\Else
\State estimating joint position vectors using Eq. (\ref{e6})
\State calculating $R_{b_1}^{b_2}$ using Eq. (\ref{xsi}) - (\ref{rsi}) and (\ref{rb1b2})
\State infering $q_{j_3}$ using Eq. (\ref{angj3}) and (\ref{qj3})
\State infering ${\tilde{q}}_{b_1}^{b_2}$ using Eq. (\ref{e22})
\State decoupling XYZ Euler angles of ${\tilde{q}}_{b_1}^{b_2}$: $\angle j_1$ $\gets$ $\theta_X$, $\angle j_2$ $\gets$ $\theta_Z$
\EndIf
\State \Return $\angle j_1$, $\angle j_2$ and $\angle j_3$
\EndWhile

\State
\Function {SenMove}{$\bm{\omega_{up}},\bm{\omega_{lo}}$}
\State $result \gets 0$
\State $\bm{[j_i^{3D}]_{det}}$ $\gets$ do 2-8 in detection windows
\State calculating $V$ using Eq. (\ref{V})
\If {$V < threshold $}
\State $result \gets 0$
\Else
\State $result \gets 1$
\EndIf
\State \Return $result$ and $\bm{[j_i^{3D}]_{det}}$
\EndFunction

\end{algorithmic}
\end{algorithm}

%\begin{table}[htbp]
%\footnotesize
%\caption{The validation of robustness towards sensor movement(Fle/Ext, Abd/Add and InR/ExR denote Flexion/Extension, Abduction/Adduction and Intro/Extra Rotation respectively)}
%\label{06}
%\centering
%\renewcommand\tabcolsep{1.0pt}
%\begin{tabular}{c|cccc}
%\toprule
%\diagbox{RMSE(deg)}{Duration} & Before Sensor Movement & During Abnormal Period & During $I_n^{det}$ & During $I_n$\\
%\midrule
%Fle/Ext & 2.54 & 204.70 & 13.97 & 3.50\\
%Abd/Add & 3.30 & 176.13 & 7.08 & 1.48\\
%InR/ExR & 1.18 & 162.14 & 7.09 & 1.97\\
%\bottomrule
%\end{tabular}
%\end{table}

\section{Experiment}
%As stated before, experiments should be constructed for analyzing errors from each step of the %algorithm and testing various metrics on human subjects. A 3-DoF gimbal was employed to isolate the %error caused by sensor-to-body alignment, thus reveals the effectiveness of reference frame %calibration and axis estimation seperately. In addition, data from experiments on human subjects %were used to estimate the accuracy of our algorithm, analyze the robustness towards sensor movement %under an extreme condition and validate the repeatability against different sets of sensor %placement.

\subsection{Validation Protocol}
\subsubsection{3-DoF Gimbal}
\begin{figure}[ht]
\centering
\subfigure[The 3-DoF gimbal]{
\includegraphics[width=4.5cm]{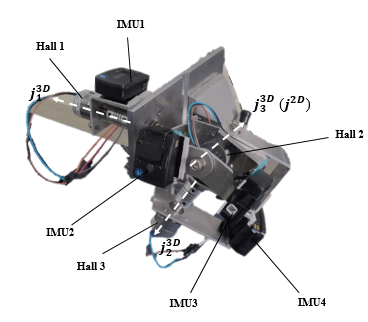}}
\subfigure[The IMU attachment on human subjects]{
\includegraphics[width=7cm]{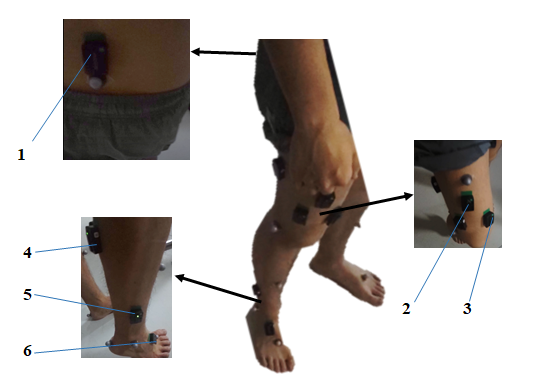}}
\caption{Experimental setup}
\label{fig:09}
\end{figure}

To validate the effectiveness of reference frame calibration separately, a gimbal consists of two segments and three rotating axes intersecting at the same point was designed to mimic a 3-DoF lower-limb joint. As shown in Fig. \ref{fig:09}, angles directly measured by Hall sensors attached to each axis by couplings, were used as reference to quantify our algorithm's performance. Four IMUs were attached to each segment of the gimbal. IMU2 and IMU3 were attached to the segments beside a 1-DoF joint whose axis was set as the main axis, while IMU1 and IMU4 were placed beside the 3-DoF joint. Due to its flat mounting surface, IMUs can be mounted with known orientation relative to body-fixed frames, which provided a reference for the estimation of axes. The gimbal was activated manually, while the largest motion was guaranteed to be around the main axis.

\subsubsection{Human subject}
For the purpose of validation on human subjects, ten healthy subjects( 23$\pm$3 years old, 175$\pm$5cm, 65$\pm$8.3kg) were enrolled in this study. To test the robustness against sensors' small oscillations caused by muscle activation, the IMUs were visually placed on major muscles of pelvis and right leg (thigh, shank, foot) with arbitrary orientations. The IMUs used in this paper were numbered as Fig. \ref{fig:09}. The subjects were asked to perform three gait modes: stair ascent, squat and level walking.Subjects were asked to firstly take four steps on level ground, ascend four steps of stair and squat three times. The sequence of these gait modes was fixed for each subject. After the multi-mode locomotion, subjects were asked to walk for 2 minutes on a treadmill after a 3-minute rest between tests to avoid the gait deviation caused by fatigue.  Optical motion capture system, Vicon, was used as gold standard for comparison purpose.

\subsection{Data Analysis}

Prior to data processing, the  raw  data  from  IMUs  were  filtered by a  low  pass  filtering method, and then method proposed  by  Feliz  et al \cite{feliz2009pedestrian} had been adopted  to  reset  the  angular  rate  to  zero  when the  angular  rate  was  under  1 $rad/s$. The bias of accelerometer and magnetometer readings were evaluated by the algorithm proposed in \cite{o2007inertial} and subtracted from the data.

During the iteration of the whole algorithm, a relatively large error might come from 1) the estimation of axes using geometric constraints, 2) the calibration of reference frames. To distinguish the sources of errors, experiments were performed on the gimbal, with IMUs being placed with proper orientations relative to gimbal axes. Hence, relative errors caused by misestimation of joint geometric constraints can be completely avoided by proper alignment.Data from  human subjects were collected and processed for validating the whole algorithm's performance on lower-limb joints, and the robustness towards different sensor placement and sensor movements.

%\begin{figure}[ht]
%\centering
%\includegraphics[width = 5cm]{10.png}
%\caption{ The schematic figure of 1-DoF joint experiment}
%\label{fig:10}
%\end{figure}

\subsubsection{Effectiveness of the reference frame calibration}
To simply analyze the effectiveness of reference frame calibration, error of estimating joint axes and position vectors should be eliminated. To do this,  $Z$ axis of IMU2 and $X$ axis of IMU3 were placed in the direction of axis $\bm{j_3^{3D}}$. In addition, structural parameters of the gimbal were used to calculate reference coordinates of joint position vectors. Thus, error caused by the deviation of reference frames can be seen as the only reason that contributes to the deviation between measured and estimated angle. The root mean square error (RMSE) between the measured 1-DoF joint angle($\theta_{measured}$) and estimated angle with ($\theta_{known}^+$) and without ($\theta_{known}^-$) calibrating reference frames was used to quantify the effectiveness of the reference frame calibration.
\begin{align}
\nonumber error_{RFC}^{1DoF}(i)&=\theta_{measured}(i)-\theta_{known}^{+/-}(i)\\
RMSE_{RFC}^{1DoF}&=\sqrt{ \frac{1}{N} \sum_{i=1}^{N}(error_{RFC}^{1DoF}(i))^2}
\end{align}

\subsubsection{Validity of axis estimation}
To validate the effectiveness of axis estimation for 3-DoF joints, an accuracy comparison needs to be made in 1-DoF axis estimation and the feedback-based iteration for estimating 3-DoF axes. Using data collected from IMU2 and IMU3, the main axis $\bm{j_1^{2D}}$, $\bm{j_2^{2D}}$, and joint position $\bm{o_1}$, $\bm{o_2}$ of the 1-DoF joint can be estimated assuming that the alignment of sensors is unknown. Then, RMSE is used to measure the performance of the algorithm, which combines estimating $\bm{j_1^{2D}}$, $\bm{j_2^{2D}}$, $\bm{o_1}$, $\bm{o_2}$ and calibrating reference frames. To present how estimates of such axes drift, the deviation between the estimated coordinates of $\bm{j_1^{2D}}$, $\bm{j_2^{2D}}$, $\bm{o_1}$, $\bm{o_2}$ and their reference coordinates known by IMUs' determined placement is used as another metic.
\begin{align}
\nonumber error_{GAE}^{1DoF}(i)^{\pm}&=\theta_{measured}^{1DoF}(i)-\theta_{estimated}^{1DoF}(i)^{\pm}\\
RMSE_{GAE}^{1DoF}(\theta^{\pm})&=\sqrt{ \frac{1}{N} \sum_{i=1}^{N}(error_{GAE}^{1DoF}(i)^{\pm})^2}
\end{align}

\begin{align}
\nonumber error(\bm{j_i^{2D}})^{\pm}&=\Vert[\bm{j^{ref}}]_{\rm}-(\bm{j_i^{2D}})^{\pm}\Vert, i=1,2\\
error(\bm{o_i})^{\pm}&=\Vert[\bm{o_i}]_{\rm}-(\bm{o_i})^{\pm}\Vert, i=1,2
\end{align}
Where $[\bm{j^{ref}}]_{\rm}$, $[\bm{o_i}]_{\rm}$ are the reference coordinates of the main axis and sensor position described in the coordinate frame of the $i$th IMU, $\pm$ denotes the estimates with and without reference frame calibration respectively.

Similarly, as a performance indicator of 3-DoF angle estimation, RMSE could also be a metric of the feedback-based iteration algorithm, which can be identified as follow:
\begin{align}
\nonumber RMSE_{GAE}^{3D}\big((\bm{j_i^{3D}})^{\pm})=\sqrt{\frac{1}{N}\sum_{i=1}^N\mu^2}\\
\mu=\theta_{measured}^{\bm{j_i^{3D}}}(i)-\theta_{estimated}^{\bm{j_i^{3D}}}(i)^{\pm}, i=1,2,3
\end{align}

Reference coordinates of the three axes, known by the placement of IMU1 and IMU4, can also provide a measurement for the estimates of such axes, given by:
\begin{align}
\nonumber error(\bm{j_i^{3D}})_{\rm}^{\pm}&=\Vert [\bm{j_i^{ref}}]_{\rm}-(\bm{j_i^{3D}})^{\pm} \Vert,\\
 i&=1, 2, 3; j=3, 4
\end{align}

Due to the varying coordinates of $\bm{j_3^{3D}}$ relative to IMU1 and IMU2, $error(\bm{j_3^{3D}})^{\pm}$ is computed at the end of each sliding window.

\subsubsection{Accuracy and Agreement of human lower-limb joint angle estimation}
To access the accuracy of the angles estimated by our algorithm, RMSE is calculated among all the lower-limb joints using estimated angles and reference angles measured by optical motion capture.
\begin{align}
\nonumber RMSE_i^j=\sqrt{\frac{1}{N}\sum_{k=1}^{N}(\theta_i^j(k)-\hat{\theta}_i^j)}\\
 i=FE, AB, IR. j=ankle, knee, hip
\end{align}
\subsubsection{Repeatability of human lower-limb joint angle estimation}
To evaluate the robustness of our algorithm against sensor placement, the repeatability is estimated by comparing RMSEs of estimated angles between different sensor placements. To do so, multiple IMUs were placed on the thigh and shank and data from different sets of IMUs were used to construct a comparative trial i.e. IMU1-IMU2, IMU1-IMU3, IMU2-IMU4, IMU3-IMU5, IMU4-IMU6, IMU5-IMU6. Then repeatability could be estimated by Bland-Altman method, which provides an interval where the errors fall with a 95\% probability \cite{bland1986statistical}.  
\subsubsection{Influence caused by different lengths of sliding windows and intervals}
To test the influence caused by different lengths of sliding window and interval, the RMSE and iteration duration were calculated with 5 sets of sliding windows and 6 sets of intervals to present their effect on accuracy and computational efficiency.

\subsubsection{Robustness against sensor movements}
An extreme condition was constructed to validate the effectiveness of sensor movement detection and safeguarded strategy. Data from IMUs besides hip joint during stair ascent were used as an example in this test. When data from IMU1 and IMU2 were used for estimating 3-DoF hip angles during upstairs, a piece of data from IMU3 was injected into data flow to replace data measured by IMU2 in the same duration. RMS errors were calculated separately to present the accuracy before detection, during detection and after detection.The algorithm in \cite{vargas2016imu}, which is proposed recently for estimating 3-D lower-limb joint angles, was employed to make a comparison with our algorithm.

\subsubsection{Accuracy under a 2-min test}
In order to demonstrate our algorithm's performance over long runs, data from the 2-min test on level walking were processed. Under this test, no sensor movement was involved. RMS errors, which quantified the accuracy,  were averaged among subjects.

%\subsubsection{Extension to a 3-DoF joint without a main axis}
%The data processing and analysis mentioned above are based on the assumption that a main axis exists during the random movements of the 3-DoF joint. One question of interest is how the algorithm performs without main axis. To answer this question, an experiment was constructed on the gimbal while the magnitudes of angles around all the three axes were ensured to be similar. For the data analysis and experimental results, please refer to the supplemental information. %In this extended experiment, data %were processed without considering real-time capability. The length of intervals were set to be as small %as possible to compensate the gap caused by processing data in a sliding window, which is shown in Fig. %\ref{fig:11}.

%\begin{figure}[h]
%\centering
%\includegraphics[width = 6cm]{11.png}
%\caption{Data windowing scheme in the extended experiment}
%\label{fig:11}
%\end{figure}

\section{Results and Discussion}

\subsection{Error Sources of 3-DoF Angle Estimation}
\begin{figure}[h]
\centering
\includegraphics[width = 8cm]{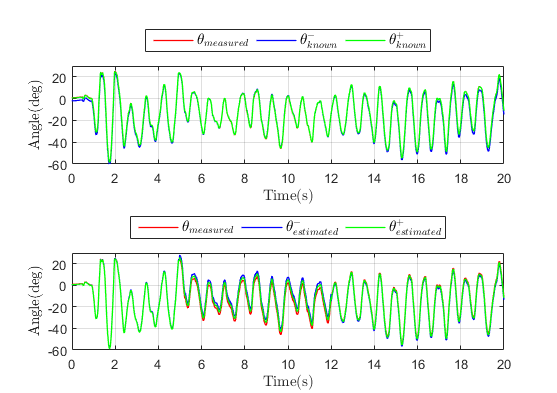}
\caption{Estimated 1-DoF angles around the axis $\bm{j^{2D}}$}
\label{fig:12}
\end{figure}

As shown in Fig. \ref{fig:12}, the curves of 1-DoF joint angle are calculated through known versus estimated (denoted by $\theta_{known}$ and $\theta_{estimated}$, respectively) coordinates of $\bm{j_1^{2D}}$, $\bm{j_2^{2D}}$, $\bm{o_1}$, $\bm{o_2}$, and with versus without (denoted by $+$ and $-$, respectively) calibrating reference frames. The root mean square errors, $RMSE_{RFC}^{1DoF}$, are 0.32 $deg$ ($+$) and 3.24 $deg$ ($-$), while $RMSE_{GAE}^{1DoF}$, are 2.06 $deg$ ($+$) and 4.17 $deg$ ($-$). The $RMSE_{RFC}^{1DoF}$ without calibration meets the error reported in \cite{yi2018estimating}.

\begin{table}[h]
\caption{Errors of 1-DoF joint axis coordinates}
\label{01}
\centering
\renewcommand\tabcolsep{3.5pt}
\begin{tabular}{ccccc}
\toprule
Name & $error(\bm{j_1^{2D}})$ & $error(\bm{j_2^{2D}})$ & $error(\bm{o_1})$ & $error(\bm{o_2})$ \\
\midrule
$+$ & 0.088 & 0.081 & 0.080 & 0.097\\
$-$ & 0.24 & 0.20 & - & -\\
\bottomrule
\end{tabular}
\end{table}

\begin{figure}[ht]
\centering
\includegraphics[width = 8.5cm]{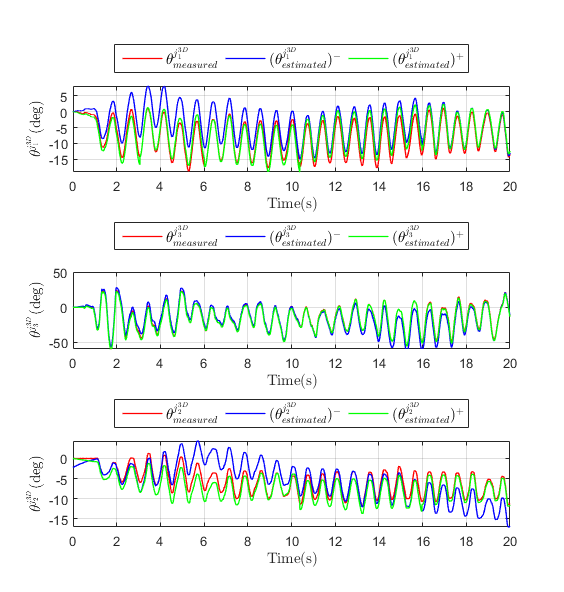}
\caption{Estimated 3-DoF angles around three joint axes $(\theta_{es}^{j_1^{3D}})^{\pm}$, $(\theta_{es}^{j_2^{3D}})^{\pm}$, $(\theta_{es}^{j_3^{3D}})^{\pm}$, where minus and plus sign denote the estimates without and with reference frame calibration respectively.}
\label{fig:13}
\end{figure}

Using data from IMU 1 and IMU 4 in the same experiment, the 3-DoF joint angle estimated by the feedback-based iteration process is presented in Fig. \ref{fig:13}. The RMS errors are listed in TABLE \ref{02}. According to the measurement of axes, the norms of estimated axes' coordinate error are shown in TABLE \ref{03}.

\subsubsection{Effectiveness of reference frame calibration}
How the  reference frame calibration affects the estimation accuracy is twofold. Firstly, because the 3-DoF geometric constraint is described in the reference frame, the calibration algorithm is embedded in the progress of estimating 3-DoF joint axes. Secondly, as shown in equation (\ref{e22}), it also involves in the construction of the body-to-body alignment, $\tilde{q}^{b_1}_{b_2}$, thus affects the estimation of 3-DoF angles.

Firstly, the reference frame calibration could contribute to a better estimation of joint axes. As shown in TABLE \ref{01} and TABLE \ref{03},the influence of whether calibrating reference frames or not affects the accuracy of estimated joint axes. It can be seen that not calibrating reference frames results in an increased error in joint axis estimation, regardless of DoFs. In addition, if a comparison is made between the curves of estimated 1-DoF angles and 3-DoF angles without calibration, as presented in Fig. \ref{fig:12} and Fig. \ref{fig:13}, it is indicated that the estimated 3-DoF angles without calibration drift a lot, which can be seen as the result of drifted estimation of joint axes in each sliding window.

Secondly, the pointwise reference frame calibration can improve the accuracy of estimated angles. By isolating the estimation of the coordinates of joint axes and joint postion vectors, as presented by $RMSE_{RFC}^{1DoF}$ and $RMSE_{GAE}^{1DoF}$ (0.32 vs. 3.24, 2.06 vs. 4.17) , the performance on the 1-DoF experiment using the gimbal  indicates even solely calibrating reference frames could make a large improvement (90\% and 50\%, respectively) on accuracy. TABLE \ref{02} presents the comparison of estimating 3-DoF angles with ($+$) versus without ($-$) pointwise reference frame calibration, which demonstrates calibrating reference frames can greatly improve the estimation accurracy of 3-DoF joints. It should be noted that the performance presented TABLE \ref{02} deteriorates drastically when cancelling reference frame calibration, which gives a proof that the cancelling of reference frame calibration could lead to an accumulation of data errors during the whole iteration process, even resulting in a wrong descending direction in LM method.

\begin{table}[h]
\caption{RMSE of estimated 3-DoF joint angles}
\label{02}
\centering
\renewcommand\tabcolsep{10.5pt}
\begin{tabular}{cccc}
\toprule
Angle & $(\theta_{es}^{j_1^{3D}})^{-/+}$ & $(\theta_{es}^{j_3^{3D}})^{-/+}$ & $(\theta_{es}^{j_2^{3D}})^{-/+}$ \\
\midrule
RMSE(deg) & 18.88/2.49 & 29.18/1.69 & 9.90/2.86\\
RMSE(\%) & 95.1/12.54 & 34.95/2.03 & 76.1/22\\
\bottomrule
\end{tabular}
\end{table}

\begin{table}[h]
\footnotesize
\caption{Errors of estimated 3-DoF axis coordinates}
\label{03}
\centering
\renewcommand\tabcolsep{8.0pt}
\begin{tabular}{cccc}
\toprule
Name & $error(\bm{j_1^{3D}})$ & $error(\bm{j_2^{3D}})$ & $error(\bm{j_3^{3D}})$ \\
\midrule
$+$ & 0.0517 & 0.0434 & 0.0089\\
$-$ & 0.4981 & 0.3702 & 0.5596\\
\bottomrule
\end{tabular}
%\begin{tablenotes}
%\item \hl{Minus and plus sign denote the estimates without and with reference frame calibration %respectively.}
%\end{tablenotes}
\end{table}

\subsubsection{Validity of axis estimation}
For 1-DoF joint angle estimation, $RMSE_{GAE}^{1DoF}$ is larger than $RMSE_{RFC}^{1DoF}$ , no matter reference frame calibrated or not, which gives an indication of to what content the 1-DoF joint axis estimation could contribute to the performance. The same data were used to analyze our algorithm's performance on the main axis $\bm{j_3^{3D}}$ of the 3-DoF joint, while the data collected simultaneously on $\bm{j_1^{3D}}$ and $\bm{j_2^{3D}}$ were added into analysis. It can be seen from TABLE \ref{02} that RMSE of $(\theta_{es}^{j_3^{3D}})^+$ is smaller than $RMSE_{GAE}^{1DoF}$ with calibrating reference frames (2.06 $deg$), which indicates that our algorithm could improve the estimation accuracy of angles around the main axis and demonstrates the vadility of decoupling a 3-DoF rotation with a main axis into rotations of three axes.

\setlength{\parskip}{0em}

\subsection{Validation on human subjects}

\subsubsection{Accuracy and agreement}
In human lower-limb joint angle estimating tests, the 3-DoF angles of hip, knee and ankle were estimated, the result of which was depicted in Fig.S3. Both lengths of sliding window ($W_n$) and interval （$I_n$）were 300 sample points for ascending and level walking while 500 sample points for squatting.The resulting RMS errors and correlation coefficients were presented in TABLE \ref{04} and TABLE S1.

\begin{table}[htbp]
%\centering
\caption{RMSE and correlation coefficients of lower-limb joint angles during level walking }
\label{04}
\begin{tabular}{c|ccc}

\toprule
\diagbox{Joint}{Direction}& Fle/Ext & Abd/Add & InR/ExR \\
\midrule
\multirow{2}{*}{Hip}  &1.72$\pm$0.89 & 0.52$\pm$0.26 & 0.74$\pm$0.44\\
& 2.5\% (0.99) & 3.9\% (0.98) & 4.0\% (0.98) \\
\cmidrule{1-4}
\multirow{2}{*}{Knee}  &1.72$\pm$0.56 & 1.82$\pm$0.83 & 0.88$\pm$0.29\\
& 3.7\% (0.99) & 4.9\% (0.98) & 2.7\% (0.96)\\
\cmidrule{1-4}
\multirow{2}{*}{Ankle}  &0.72$\pm$0.20 & 0.44$\pm$0.13 & 1.92$\pm$1.09\\
& 2.5\% (0.97) & 8.1\% (0.94) & 4.0\% (0.97)\\
\bottomrule
\end{tabular}
\begin{tablenotes}
\item The decimals in brackets denote correlation coefficients of each joint. Numbers after $\pm$ denote the stantard deviation of RMSE among subjects. For performance during stair ascent and squat, please refer to the Supplementary information.
\end{tablenotes}
\end{table}

%\begin{table*}[htbp]
%\caption{RMS Errors of lower-limb joint angles(Fle/Ext, Abd/Add and InR/ExR denote Flexion/Extension, %Abduction/Adduction and Intro/Extra Rotation %respectively)}
%\label{04}
%\centering
%\begin{tabular}{c|ccc|ccc}
%\toprule
%\multirow{2}{*}{} & \multicolumn{3}{c}{Ascend} & \multicolumn{3}{c}{Squat}\\
%\cmidrule{2-7}
%&Fle/Ext & Abd/Add & InR/ExR & Fle/Ext & Abd/Add & InR/ExR\\
%\midrule
%\multirow{2}{*}{Hip} & $RMSE_{FE}^{Hip}$ & $RMSE_{AB}^{Hip}$ & $RMSE_{IR}^{Hip}$ & $RMSE_{FE}^{Hip}$ &  %$RMSE_{AB}^{Hip}$ & $RMSE_{IR}^{Hip}$\\
%&1.78$\pm$0.73/3.2\% & 2.57$\pm$1.27/16.1\% & 1.53$\pm$0.59/8.2\% & 2.18$\pm$1.45/2.3\% & %1.83$\pm$0.73/21.4\% & 1.28$\pm$0.21/7.3\%\\
%\cmidrule{1-7}
%\multirow{2}{*}{Knee} & $RMSE_{FE}^{Knee}$ & $RMSE_{AB}^{Knee}$ & $RMSE_{IR}^{Knee}$ & $RMSE_{FE}^{Knee}$ %&  $RMSE_{AB}^{Knee}$ & $RMSE_{IR}^{Knee}$\\
%&2.87$\pm$0.82/3.2\% & 4.94$\pm$0.19/27.8\% & 2.47$\pm$1.19/11.3\% & 2.98$\pm$1.07/3.2\% & %3.15$\pm$0.91/16.3\%& 2.67$\pm$0.87/9.6\%\\
%\cmidrule{1-7}
%\multirow{2}{*}{Ankle} & $RMSE_{FE}^{Ankle}$ & $RMSE_{AB}^{Ankle}$ & $RMSE_{IR}^{Ankle}$ & %$RMSE_{FE}^{Ankle}$ &  $RMSE_{AB}^{Ankle}$ & %$RMSE_{IR}^{Ankle}$\\
%&1.24$\pm$0.0.41/6.0\% & 1.44$\pm$0.17/15.1\% & 0.66$\pm$0.13/12.8\% & 3.08$\pm$1.78/9.1\% & %2.21$\pm$0.54/38.3\% & 0.39$\pm$0.77/6.6\%\\
%\bottomrule
%\end{tabular}
%\end{table*}

%\soulregister\cite{IEEEexample:Seel2014IMU}

 In addition to testing our algorithm's performance on level walking, we constructed the validation on much worse conditions with larger acceleration and more severe skin artificial movements, which were stair ascending and squatting. During such tasks, we can still obtain better accuracy in angle around the main axis compared with another self-alignment method reported in \cite{seel2014imu} during level walking, while angles in the other two directions were estimated simultaneously with relatively good accuracy. 

\subsubsection{Repeatability}
Fig. S4 clearly presents the repeatability toward different placement of sensors using the statistics method proposed in \cite{bland1986statistical}. As shown in Fig. S4, over 95\% difference between two IMU sets fell in the interval of mean$\pm$SD for all the subfigures. Through the result we have obtained, we can conclude that our algorithm is sensor-placement free during the axes' estimation, in other words, the progress of sensor self-alignment. 

\subsubsection{Influence caused by different lengths of sliding windows and intervals}

The overall  effect  resulted  from  different  lengths of sliding window and interval has been verified by an error metric, given by:
\begin{align}
\label{e30}
\nonumber f_n(W_n,I_n)=\sum_j b_j \frac{\mu}{a_1+a_2+a_3},j=hip,knee,ankle\\
\mu=\frac{a_1}{\mu_{FE}^j}RMSE_{FE}^j+\frac{a_2}{\mu_{AB}^j}RMSE_{AB}^j+\frac{a_3}{\mu_{IR}^j}RMSE_{IR}^j
\end{align}

Where $f_n(W_n,I_n)$ denoted a weighted sum of RMS errors of every joint and every direction, $\mu_i^j(i=FE,AB,IR)$ was the correlation coefficients of each motion and joint. Due to the same importance of each joint, $b_j$ was set to 1. We set $a_1=2$ and $a_2=a_3=1$, considering that more attention was paid to Flexion/Extension during most gait research. For each $W_n$ and $I_n$, Fig. S5 was presented to show the variety of performance. For the sake of real time application, the sliding window should be greater than or equal to the interval $I_n$.

%\begin{figure}[htbp]
%\centering

%\subfigure[Walking]{
%\begin{varwidth}[t]{0.2\textwidth}
%\vspace{0pt}

%\includegraphics[width = 3.6cm]{15c.png}

%\end{varwidth}
%}
%\qquad
%\subfigure[Stait ascent]{
%\begin{varwidth}[t]{0.2\textwidth}
%\vspace{0pt}

%\includegraphics[width = 3.6cm]{15a.png}

%\end{varwidth}
%}
%\qquad
%\subfigure[Squat]{
%\begin{varwidth}[t]{0.2\textwidth}
%\vspace{0pt}

%\includegraphics[width = 3.6cm]{15b.png}

%\end{varwidth}
%}
%\caption{The comprehensive error metric of the overall effect caused by different sliding% %windows and intervals.}
%\label{fig:15}
%\end{figure}

%\begin{figure*}[ht]
%\centering
%\subfigure[Stair ascent]{
%\includegraphics[width=0.48\textwidth, height = 7cm]{17a.png}}
%\subfigure[Squat]{
%\includegraphics[width=0.48\textwidth, height = 7cm]{17b.png}}
%\caption{The repeatibility represented by the difference-average estimates.}
%\label{fig:17}
%\end{figure*}

It  can be indicated from Fig. S5 that the metric calculated by equation (\ref{e30}) gently varies with the changing lengths of sliding window and interval. Neither smaller nor greater sliding window length were included in this validation because algorithm with a smaller sliding window, although convergent, converges to a saddle point which was far away from the optimal solution. This could be given a side proof, shown in TABLE \ref{06}, by the RMSE of sensor movement during $I_n^{det}$, in which $(\bm{j_i^{3D}})_{det}$ were estimated by just 50 measurements. These RMS errors were obviously larger than those in TABLE S1. And the main axis didn't stay still relative to IMUs. In contrast, its coordinates in sensor-fixed frames were slowly varying, the speed of which depended on the sort of moving tasks and the locomotion features of subjects. So a greater sliding window cannot improve the accuracy either. Selection of interval length $(I_n)$ was related to the length of sliding window $(W_n)$. During every interval $I_n$, angles were computed according to the axes estimated in last sliding window.  $I_n$ should be equal or greater than $W_n$, while a too great $I_n$ would reduce the accuracy. So an interval $[W_n,800]$ was considered by compromising the real time application and accuracy requirements.

%\begin{table}[htbp]
%\caption{Correlation Coefficient(Fle/Ext, Abd/Add and InR/ExR denote Flexion/Extension, Abduction/Adduction and Intro/Extra Rotation respectively)}
%\label{05}
%\centering
%\renewcommand\tabcolsep{4.0pt}
%\renewcommand\arraystretch{1.5}
%\begin{tabular}{c|ccc|ccc}
%\Xhline{0.8pt}
%& \multicolumn{3}{c}{Ascend} & \multicolumn{3}{c}{Squat}\\
%%\Xcline{2-7}{0.4pt}
%&Fle/Ext & Abd/Add & InR/ExR & Fle/Ext & Abd/Add & InR/ExR\\
%\Xhline{0.4pt}
%Hip & 0.99 & 0.98 & 0.98 & 0.99 & 0.96 &0.98\\
%\Xhline{0.4pt}
%Knee & 0.99 & 0.78 & 0.94 & 0.99 & 0.90 & 0.96\\
%\Xhline{0.4pt}
%Ankle & 0.96 & 0.92 & 0.86 & 0.97 & 0.98 & 0.93\\
%\Xhline{0.8pt}
%\end{tabular}
%\end{table}

\begin{figure}[h]
\centering
\includegraphics[width = 5cm, height=4cm]{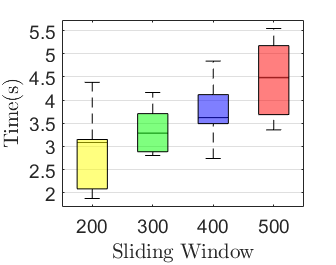}
\caption{The boxplot of computing time versus sliding window.}
\label{fig:16}
\end{figure} 

Fig. \ref{fig:16} presented the computing time of 6 iterations with different lengths of sliding window. It is shown that the computing time was increasing with the lengthening sliding window.

\subsubsection{Accuracy under the 2-min test}
As shown in TABLE \ref{T2-min} and Fig. S6, the RMSE were presented after the five subjects performed the 2-minute test. Compared with the performance of the short-run test (shown in TABLE S1), our algorithm on the 2-min test presented a similar mean and a slightly larger standard deviation among subjects. This performance can demonstrate our algorithm's stability under a long-run test.

\begin{table}[h]
\caption{RMSE of the 2-min test}
\label{T2-min}
\centering
\begin{tabular}{cccc}
\toprule
\diagbox{Joint}{Direction} &Fle/Ext & Abd/Add &InR/ExR \\
\midrule
Hip & 1.69$\pm$ 0.90  & 0.52$\pm$ 0.33 & 0.79 $\pm$ 0.43\\
Knee & 1.74 $\pm$ 0.77 & 1.77 $\pm$ 0.83 & 0.83 $\pm$ 0.43\\
Ankle & 1.25 $\pm$ 0.90 & 0.39 $\pm$ 0.71 & 0.73 $\pm$ 0.81\\
\bottomrule
\end{tabular}
\end{table}

\subsection{The robustness towards sensor movement}
\begin{figure}[h]
\centering
\includegraphics[width = 8.5cm]{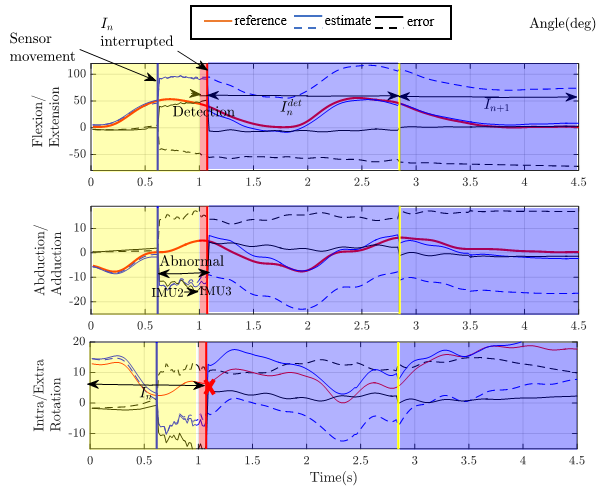}
\caption{The validation of robustness towards sensor movement. Solid lines denote the estimation result of our algorithm, while dashed lines denote the results estimated by the algorithm proposed by \cite{vargas2016imu}}
\label{fig:18}
\end{figure}

The robustness towards sensor movements was validated in Fig. \ref{fig:18}. The RMS errors before sensor movement, during abnormal period, during $I_n^{det}$ and during ${I_n}$ were presented in TABLE \ref{06}.

\begin{table}[h]
\footnotesize
\caption{The validation of robustness towards sensor movement}
\label{06}
\centering

\begin{tabular}{c|c|c|c|c}
\toprule
\diagbox{RMSE(deg)}{Duration} & {BSM} & DAP &  $I_n^{det}$ &  $I_n$\\
\midrule
\multirow{2}{*}{Fle/Ext} & 2.54 & 43.60 & 13.97 & 3.50\\

& (2.64) & (45.19) & (57.54) & (47.90)\\
\hline
\multirow{2}{*}{Abd/Add} & 1.23 & 15.50 & 1.93 & 1.19\\

& (0.64) & (15.19) & (14.23) & (16.80)\\
\hline
\multirow{2}{*}{InR/ExR} & 1.89 & 11.89 & 4.26 & 2.12\\
& (1.21) & (11.39) & (11.73) & (13.02)\\

\bottomrule
\end{tabular}
\begin{tablenotes}
\item Fle/Ext, Abd/Add and InR/ExR denote Flexion/Extension, Abduction/

Adduction and Intro/Extra Rotation respectively.BMS denotes before 

sensor movement, DAM denotes during abnormal period.The numbers inside brackets are RMSE of the algorithm in \cite{vargas2016imu}.
\end{tablenotes}
\end{table}

 It is shown in  Fig. \ref{fig:18} and TABLE \ref{06}, during abnormal period after sensor movement, a large deviation was induced by replacing IMUs. After the variation metrics $V$ was detected to exceed the threshold we set, the normal interval $I_n$ was interrupted (denoted by the vertical red line). During the temporal interval $I_n^{det}$, the estimates in detection widow (denoted by the red area) were used to calculate 3-DoF angles. As we can see in TABLE \ref{06}, the RMSE in this duration reduced a lot but were still larger than RMSE in next normal interval $I_{n+1}$. Comparing the results estimated by our algorithm and the algorithm in \cite{vargas2016imu}, it is indicated that our algorithm presented strong robustness towards sensor movement, although suffered from a slightly lower accuracy during BMS.

\section{Conclusion}
This study demonstrates an initial attempt to develop and evaluate a sensor-movement-free 3-DoF joint angle estimation algorithm for lower-limb joints. A pointwise  reference  frame  calibration  method  and  a  feedback-based iteration progress are presented. In the experiments  with the  3-DoF  gimbal,  errors of  sensor-to-body  alignment  and  reference  frame  calibration have been estimated separately. On human subjects, the robustness against sensor placement and sensor movement are validated respectively by the repeatability of different sets of IMUs and a  set  of  RMSE  during  detecting  and  compensating  sensor movement.  The  results of  this  pilot  study  have shown that  the feedback-based  iteration  design  is  visible  for  estimating  3-DoF  joint  axes  and  the  novel  reference  frame  calibration algorithm  is able to improve  accuracy  and  promote  the  convergence  of  the whole  estimation  algorithm.  Robustness  against sensor  placement  and  movement has been demonstrated and real-time  application has been guaranteed  by  validating  the computing time. However, continuing efforts are still required to  further  improve the  robustness  towards  various  lengths  of sliding window and interval, and a more computing efficient algorithm need to  be  adopted  if  the  algorithm  is  applied for the estimation of 3-DoF joint angle without main axis.

\appendices

%\section{}

\section*{Acknowledgment}

This research was supported by the Robotics and Rehabilitation Lab, Harbin Institute of Technology. Acknowledgment to B$\&$R Intelligence Research Exchange Center.

\bibliographystyle{IEEEtran}

\bibliography{paper02}
\end{document}